\documentclass{andromedaone}       

\def\be{\begin{equation}}
\def\ee{\end{equation}}
\def\bea{\begin{eqnarray}}
\def\eea{\end{eqnarray}}

\begin{document}

\title{Speculation about the black hole final state:\\ resolving singularity by quantum gravity\footnote{Proceedings of \textit{Beyond Standard Model: From Theory to Experiment}, the 29th of March, 2021, Zewail City of Science and Technology, Giza, Egypt}}

\author{Dong-han Yeom\auno{1}}
\address{$^1$Department of Physics Education, Pusan National University, Busan 46241, Republic of Korea}
\address{$^2$Research Center for Dielectric and Advanced Matter Physics, Pusan National University, Busan 46241, Republic of Korea}

\begin{abstract}
The interior of the black hole can be described by anisotropic cosmology. By quantizing the metric function, we can obtain the Wheeler-DeWitt equation for inside the horizon. In order to interpret the wave function consistently, one needs to impose a boundary condition. In this paper, we introduce a prescription for the Euclidean analytic continuation inside the horizon and the corresponding wave function solution.
\end{abstract}

\maketitle


\section{Introduction}

Understanding the inside structure of the black hole in the light of quantum gravity is the fundamental but unresolved problem of modern theoretical physics. This topic is related to the resolution of the singularity \cite{Bojowald:2018xxu} as well as the information loss problem of black holes \cite{Hawking:1976ra}, where we need to figure out the true answer for the question \cite{Yeom:2008qw}. There are several approaches toward the quantum gravity, but the most constructive approach to resolve the singularity inside the black hole horizon is to follow the canonical approach, i.e., to solve the Wheeler-DeWitt equation \cite{DeWitt:1967yk}.

Regarding this approach, there are several important physical questions.
\begin{itemize}
\item[--1.] What is the \textit{boundary condition} of the wave function for the inside the horizon? Is there any unique or trivial way to provide the boundary condition?
\item[--2.] What is the \textit{fate of the infalling observer}? Will the classical geometry be extended? If not, what is the destiny of the infalling information?
\item[--3.] What is the \textit{global description including both of inside and outside the horizon}? In addition to this, what is the implication to the information loss problem?
\end{itemize}
Although the answers to all these questions are beyond the scope of this paper, we will mainly focus on the first issue.

We first provide the generic framework to describe inside the horizon using the canonical formalism \cite{Cavaglia:1994yc}. In principle, there can be several ways to choose the wave function, but one may ask what is the ground state of the wave function. Of course, there is no well-defined notion of the ground state in quantum gravity, but we can remind the wisdom from the Hartle-Hawking wave function \cite{Hartle:1983ai}. If one can provide the Euclidean future boundary condition for the inside the black hole, this will mimic the idea of the no-boundary wave function; here, Horowitz and Maldacena further speculated that this final state can provide the projection of the wave function \cite{Horowitz:2003he}. Can we realize and justify, or criticize these ideas \cite{Yeom:2008nt}?

\section{Wave function inside the horizon}
\label{sec:wav}

Let us first choose the metric ansatz for inside the horizon as follows \cite{Bouhmadi-Lopez:2019kkt}:
\begin{eqnarray}
ds^{2} = - N^{2}(t) dt^{2} + a^{2}(t) dr^{2} + \frac{r_{s}^{2} b^{2}(t)}{a^{2}(t)} d\Omega^{2},
\end{eqnarray}
where $N(t)$ is the lapse function, $r_{s}$ is the Schwarzschild radius, and $a(t)$ and $b(t)$ denote two canonical variables. Here, the classical Schwarzschilid solution correspond to the following relation:
\begin{eqnarray}
\frac{1}{b} = a + \frac{1}{a},
\end{eqnarray}
where this relation is invariant up to the choice of the lapse function $N(t)$. From this metric ansatz, one can derive the Wheeler-DeWitt equation:
\begin{eqnarray}
\left( \frac{\partial^{2}}{\partial X^{2}} - \frac{\partial^{2}}{\partial Y^{2}} + 4 r_{s}^{2} e^{2Y} \right) \Psi \left(X,Y\right) = 0,
\end{eqnarray}
where $X = \ln a$ and $Y = \ln b$.

In order to solve the partial differential equation, one can first introduce the separation of variables, say $\Psi = \phi(X) \psi(Y)$, where
\begin{eqnarray}
\left(\frac{d^{2}}{dX^{2}} + k^{2}\right) \phi(X) &=& 0, \\
\left(\frac{d^{2}}{dY^{2}} - 4 r_{s}^{2} e^{2Y} + k^{2}\right) \psi(Y) &=& 0,
\end{eqnarray}
where $k^{2}$ is the separation constant. Interestingly, this is equivalent to the two-dimensional Schr\"odinger equation of quantum mechanics, where $X$ is a time-like direction and $Y$ is a space-like direction with the potential barrier $\sim e^{2Y}$.

Following the usual wisdom of quantum mechanics, we can obtain the following observations:
\begin{itemize}
\item[--1.] It is reasonable to interpret $k^{2} > 0$ as the energy-like eigenvalue.
\item[--2.] $X$ and $Y$ domains are not bounded; hence, $k^{2}$ is not quantized.
\item[--3.] There is only the potential barrier along the $Y$-direction, and the barrier is divergent as $Y$ goes to infinity. Hence, if there is an incoming mode along the $+Y$-direction, then there must be an outgoing mode along the $-Y$-direction (unless there exist divergent contributions at the $Y > 0$ region).
\item[--4.] The classical observer will follow the Ehrenfest theorem, i.e., the peak of the wave function should correspond to the classical solution. Therefore, at the horizon, we need to impose the boundary condition that the wave function has a peak at the classical solution.
\end{itemize}
By imposing these properties, we obtain the following form of the solution without loss of generality \cite{Bouhmadi-Lopez:2019kkt}:
\begin{eqnarray}
\Psi(X,Y) = \int_{-\infty}^{\infty} f(k) e^{-ikX} K_{ik}\left(2r_{s} e^{Y}\right) dk,
\end{eqnarray}
where $K_{ik}$ is the hyperbolic Bessel function. If we choose
\begin{eqnarray}
f(k) = \frac{2A e^{-\sigma^{2}k^{2}/2}}{\Gamma(-ik)r_{s}^{ik}},
\end{eqnarray}
then the classical solution is located at the Gaussian peak of the wave function at the event horizon ($X, Y \rightarrow - \infty$), where $\sigma$ is the standard deviation and $A$ is the normalization constant.

\section{Euclidean analytic continuation}

Since $X$ is the time-like variable in the superspace, one may ask what happens if we Wick-rotate $X = -ix$. In usual quantum mechanics, along the Euclidean time, all excited modes exponentially decay, and hence the Euclidean Wick-rotation helps to choose the ground state wave function. However, in our case, after the Wick-rotation, $k < 0$ modes exponentially blow up. Therefore, in order to provide a consistent Wick-rotation, we need to choose only $k > 0$ modes, i.e., if $X < 0$,
\begin{eqnarray}
\Psi(X,Y) = \int_{0}^{\infty} f(k) e^{-ikX} K_{ik}\left(2r_{s} e^{Y}\right) dk, \label{eq:1}
\end{eqnarray}
and after the Wick-rotation at $X = -ix$, we obtain
\begin{eqnarray}
\Psi(x,Y) = \int_{0}^{\infty} f(k) e^{-kx} K_{ik}\left(2r_{s} e^{Y}\right) dk. \label{eq:2}
\end{eqnarray}

\begin{figure}
\begin{center}
\includegraphics[scale=0.6]{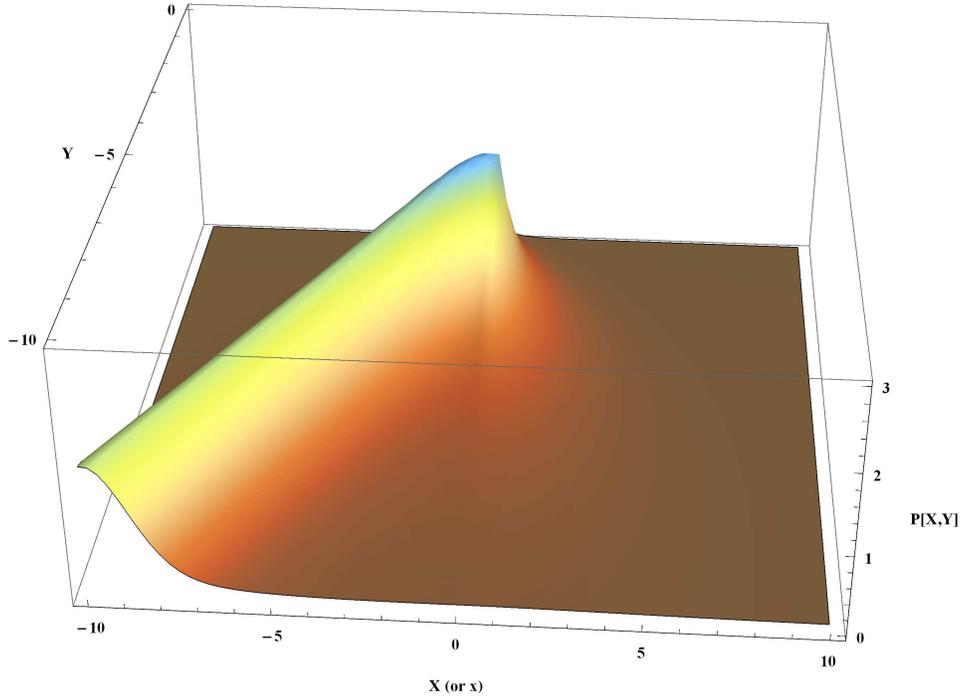}
\caption{\label{fig:res}The modulus square of the wave function $P[X,Y] = |\Psi|^{2}$, where $X < 0$ is Eq.~(\ref{eq:1}), $x > 0$ is Eq.~(\ref{eq:2}), and $\sigma = r_{s} = A = 1$.}
\end{center}
\end{figure}

A numerical example is reported in Fig.~\ref{fig:res}. Even though we choose $k > 0$ modes, at the event horizon, still the Ehrenfest theorem will be satisfied (left half of Fig.~\ref{fig:res}). However, near the bouncing point $X \sim 0$, the bias from the classical trajectory becomes significant. After the Wick-rotation, as expected, the wave function quickly decays to zero (right half of Fig.~\ref{fig:res}).

Then what will be the corresponding Euclidean on-shell geometry for this proposed Euclidean wave function (see also speculations in \cite{Chen:2016ask})? Indeed, this question has no meaning, because there is no trivial steepest-descent in the Euclidean part, as we can see in Fig.~\ref{fig:res}. However, we can check whether we can regard the $x > 0$ part as the genuine Euclidean metric or not. The clue might be related to the Wick-rotation $X = - ix$. The metric ansatz becomes
\begin{eqnarray}
ds^{2} = - N^{2}(t) dt^{2} + e^{-2ix(t)} dr^{2} + r_{s}^{2} e^{2Y(t)}e^{+2ix(t)} d\Omega^{2}.
\end{eqnarray}
In the first glimpse, this metric looks a little bit strange because the metric is entirely complex-valued. However, as $x$ increases, the wave function rapidly decays to zero. So, if we choose a reasonable real-valued Euclidean manifold at the sufficiently large $x$, this is enough. If we choose $x = n\pi/2$, the metric becomes
\begin{eqnarray}
ds^{2} = - \left( dT^{2} + dr^{2} + r_{s}^{2} e^{2Y(T)} d\Omega^{2} \right),
\end{eqnarray}
and hence, the signature is Euclidean (by choosing a proper $N(t)$, we can define $T$).

\section{Conclusion}

In this paper, we sketched the idea to impose the Euclidean boundary condition of the Wheeler-DeWitt wave function for the inside the horizon. The motivation was originated from the Hartle-Hawking proposal, but in the black hole case, the corresponding Euclidean geometry is less clear. However, there are several important physical meanings.
\begin{itemize}
\item[--] First, the wave function Eq.~(\ref{eq:2}) satisfies several physical criteria that we discussed in Sec.~\ref{sec:wav}.
\item[--] Second, as $x$ increases, the high-$k$ mode rapidly decreased. If we interpret that $k$ is the analog of the energy, then this Euclidean boundary condition selects the lowest energy state, or the ground state.
\item[--] Third, by choosing sufficiently large $x$, one can find a real-valued geometry that has zero probability but has Euclidean signatures. From this starting point, one can construct a reasonable complex contour from the Euclidean geometry to the Lorentzian geometry, as we usually did in the Euclidean path-integral approach.
\end{itemize}
Therefore, one may speculate that Eq.~(\ref{eq:2}) is the ground state wave function as a solution of the Wheeler-DeWitt equation, although we need further justifications.

One may further ask whether this wave function can explain the projection of the wave function as Horowitz and Maldacena proposed. However, this seems less probable, because the wave function depends on the incoming data or the physical information of the event horizon. However, this Euclidean boundary condition may have important physical meaning to the information loss paradox. The wave function of the black hole asymptotically approaches to zero in the superspace. Therefore, if there exists the singularity as well as the inside structure of the semi-classical history, the overall probability should approach to zero eventually; hence, the only histories without singularity will be dominated eventually \cite{Sasaki:2014spa}. This boundary condition seems to be one another realization of the DeWitt boundary condition in some sense \cite{Bouhmadi-Lopez:2019kkt}. There can be further interesting speculations, but we leave them for future investigations.

\section*{Acknowledgements}
DY is supported by the National Research Foundation of Korea (Grant no.:2021R1C1C1008622).

\bibliographystyle{unsrt}

\end{document}